\documentclass[pra, twocolumn, a4paper, superscriptaddress, secnumarabic]{revtex4-2}

\usepackage[T1]{fontenc}
\usepackage[utf8]{inputenc}

\usepackage{graphicx}
\usepackage{amsmath}
\usepackage{amsfonts}
\usepackage{amssymb}
\usepackage{hyperref}
\usepackage{float}
\usepackage{nicematrix}
\usepackage{enumitem}

\usepackage[version=4, arrows=pgf-filled]{mhchem}
\usepackage{bbm}
\usepackage{tikz}
\usepackage{diagbox}
\usepackage{makecell}
\graphicspath{{./Figures/}}

\newcommand{\X}{X}

\renewcommand{\d}{\mathrm{d}}

\newcommand{\boldx}{\boldsymbol{c}}

\newcommand{\W}{\mathcal{W}}
\newcommand{\M}{\mathbb{M}}

\renewcommand{\S}{\mathbb{S}}

\begin{document}

\title{Stability conditions of chemical networks in a linear framework}

\author{Armand Despons}
\email{armand.despons@pm.me}
\affiliation{Gulliver Laboratory, UMR CNRS 7083, PSL Research University, ESPCI, Paris F-75231, France}

\author{Jérémie Unterberger} 
\affiliation{Institut Elie Cartan, Laboratoire Associé CNRS UMR 7502, Université de Lorraine, B.P. 239, 54506 Vand\oe vre-lès-Nancy, France}

\author{David Lacoste}
\affiliation{Gulliver Laboratory, UMR CNRS 7083, PSL Research University, ESPCI, Paris F-75231, France}

\date{\today}
\begin{abstract}
Autocatalytic chemical reaction networks can collectively replicate or maintain their constituents despite degradation reactions only above a certain threshold, which we refer to as the decay threshold. 
When the chemical network has a Jacobian matrix with the Metzler property, we leverage analytical methods developed for Markov processes to show that the decay threshold can be calculated by solving a linear problem, instead of the standard eigenvalue problem. 
We explore how this decay threshold depends on the network parameters, such as its size, the directionality of the reactions (reversible or irreversible), and its connectivity, then we deduce design principles from this that might be relevant to research on the Origin of Life.
\end{abstract}

\maketitle


\section{Introduction}

Mechanisms that are able to explain how complexity can increase in chemistry are of great interest for research on the origins of life \cite{Robinson2022}. 
The appearance of large and functional molecules, which could serve as catalysts is likely to be an important step in this increase of chemical complexity. To compensate for the potential loss of catalysts, due to unwanted side reactions or unavoidable degradation mechanisms, one generally invokes autocatalytic reaction networks or cycles \cite{Xavier2020, Lancet2018, Schuster2019, Cornish2020, Ameta2021,orgel_2008}. 
A relevant question is therefore to determine the threshold above which autocatalytic reaction networks typically overcome degradation and dilution effects, which we call the decay threshold.

Assuming autocatalytic species form a single cycle and participate in unimolecular irreversible reactions, King derived such a simple criterion for the onset of growth in such a network \cite{King1982}. 
That criterion was recognized as key for understanding molecular replication in the origin of life, because it plays a similar role as the error threshold introduced by Eigen for the maintenance of information \cite{eigen2012hypercycle,szathmary_origin_2006}. While some generalizations of the King criterion to reversible reactions and to associations of coupled autocatalytic cycles have been considered before \cite{Sarkar2019,Blokhuis2020}, there is no explicit criterion for an arbitrary complex chemical network. 
On the numerical side, the situation is of course, much better: using simulations of autocatalytic sets (ACS), the threshold for the onset of growth in ACS, and interaction effects have also been studied in detail. For instance, it was observed that this threshold for onset of growth is typically lower in nested ACS due to reinforcement effects occurring between ACS \cite{giri_origin_2012}. These reinforcement effects could be potentially important for research on the origin of life, since they could offer a route to produce large molecules in the absence of enzymes.

In a recent work \cite{Blokhuis2020}, a stoichiometric definition of autocatalysis was introduced (see also \cite{Andersen2020} for a discussion of alternative approaches), which leads to a method of identification of autocatalysis within large and complex chemical networks that does not require any knowledge of kinetics \cite{gagrani_polyhedral_2024,Peng2020}. In this work, autocatalytic networks were also shown to contain minimal subnetworks called autocatalytic cores, which can only exist in five types.
These cores have the remarkable property that their stoichiometric matrix is a Metzler matrix. 
The relation between stoichiometric and dynamical definitions of autocatalysis is an active topic explored in several works \cite{lacoste_2025}. Let us mention some of them briefly here:

A first work proved that all chemical reaction networks with linearizable rates (i.e. rates that grow linearly with the concentrations in a neighborhood of zero, including mass-action and Michaelis-Menten) have a Metzler Jacobian in a dilute regime, i.e. for small enough concentrations \cite{unterberger_stoechiometric_2022}. 
In this regime, one can prove rigorously an equivalence between the stoichiometric and dynamical definitions of autocatalysis in a strongly connected network.

Beyond the dilute regime, it was shown that for a stoichiometrically autocatalytic network in the sense of \cite{Blokhuis2020}, there exists a choice of kinetic parameters for which the dynamics can be entirely dominated by the autocatalytic part, when the kinetics is parameter-rich  \cite{stadler2024}.
Then, Kosc et al. showed that autocatalytic cores can be realized dynamically in a CRN, which means that a set of chemical species at positive concentrations and a set of thermodynamically consistent reactions satisfying mass-action law kinetics can be found for all of them \cite{Kosc2025}. In addition, the fluxes within the cores share the same orientation.  
Taken together, these two works imply that autocatalytic cores represent a part of the network that can be effectively described by irreversible reactions and linear kinetics. This means that for any chemical network which is expected to grow from the stoichiometric point of view, there should exist at least one subnetwork, with a Metzler Jacobian.

Due to the linearity property of cores, typical non-linear features of chemical reaction networks (CRNs) such as multi-stability should be absent at the level of individual autocatalytic cores, a result that can indeed be proven rigorously \cite{Nandan2025}.

Inspired by these specific properties of autocatalytic cores, we consider in this paper a class of linear chemical reaction networks, for which the Jacobian matrix is Metzler \cite{Smith_1986}. This is a central assumption for this work, but it does not need to concern a large network, it could hold at the level of a subnetwork within a larger network.
Metzler matrices have wide applications in economics, biology, and in control theory \cite{Briat2017}.
In the literature, a Metzler matrix is a square matrix with non-negative off-diagonal elements, with no constraint on the diagonal  \cite{meyer2023}. 
We prove that the elements on the diagonal of the Jacobian are generally negative under mild conditions for CRNs, and detail conditions on the kinetics for the off-diagonal elements to be positive. CRNs with a Metzler Jacobian include, but are not limited to :
(i) catalytic \cite{Hideshi2024} and autocatalytic reaction networks with linear kinetics and degradation reactions \cite{unterberger_stoechiometric_2022} or (ii) autocatalytic cores with irreversible reactions. 

Our paper is organized as follows.
In the next section, we show that this central Metzler property translates into a framework that we build on the theory of continuous-time Markov chains. 
We illustrate the method on several networks of simple topology, and we use it to compare the decay thresholds for type I and type II autocatalytic cycles. 
Then, we study the effect of mutual reinforcement of autocatalytic cycles through side branches or through coupling via shared chemical species and reactions. 
This effect has similarities with the reinforcement coupling observed in \cite{giri_origin_2012}, although there are also important differences since our approach is linear, while non-linear kinetics is an essential ingredient in that work. We conclude with a discussion of the possible relevance of these findings to the Origin of Life research.

\section{General criterion for stability}

\subsection{Introduction to CRNs} 

Let us consider a chemical reaction network, defined by a certain set of chemical species and reactions. To be fully general, let us also assume that reactions are reversible, although the conditions stated above of linearity and Metzler properties of the CRN are more easily satisfied when reactions are assumed to be irreversible. 
A chemical reaction $\rho$ in the network may be written as 
\begin{equation}
\label{eq:general_transformation}
\sum_i S_{i \rho}^+ \ x_i \ce{<=>}\sum_i S_{i \rho}^- \ x_i,
\end{equation}
where $S^+_{i \rho}$ (resp. $S^-_{i \rho}$) is the stoichiometric coefficient associated with species $x_i$ for reaction $\rho$ on the reactant (resp. product) side.
The net stoichiometric change of species $x_i$ for reaction $\rho$ is then $S_{i \rho} = S^-_{i \rho} - S^+_{i \rho}$.
By gathering these coefficients, we define the stoichiometric matrix, $\S = \left\lbrace S_{i \rho} \right\rbrace$, which encodes the topology of the underlying reaction network.
The rate at which reaction Eq.~\eqref{eq:general_transformation} proceeds is $f_\rho$, a function of the species concentrations as well as other kinetic parameters.
The deterministic kinetic rate equations take the form:
\begin{equation}
\label{eq:kinetic_eqn_genral}
\d_t \boldsymbol{c} = \S \cdot \boldsymbol{f}(\boldsymbol{c}),
\end{equation}
where $\boldsymbol{c} = \left\lbrace c_i \right\rbrace$, where $c_i$ is the concentration of species $x_i$, and $\boldsymbol{f} = \left\lbrace f_\rho \right\rbrace$ is the flux vector. This flux vector can itself be split into a contribution from forward and backward reactions
$f_\rho=f_{+ \rho} - f_{- \rho}$.

One particular choice of rate laws, which can be made thermodynamically consistent and encompasses many known rate laws \cite{Wolf_2010} takes the form
\begin{equation}
    \label{eq:bio_rate}
    f_\rho (\boldsymbol{c} ) =  
        \dfrac{k_\rho^+ \prod_i \left( c_i \right)^{S^+_{i \rho}} - k_\rho^- \prod_i \left( c_i \right)^{S^-_{i \rho} }
         }{
            D_\rho (\boldsymbol{c} )
        },
    \end{equation}
where $k_\rho^\pm$ represents the kinetic constant for the forward (resp. backward) reactions; and $D_\rho (\boldsymbol{c} )$ is a polynomial such that, for each of its monomials in terms of the reactant $x_i$, the maximal degree is $S^+_{i \rho}$, and for each of its monomial in terms of the product $x_j$, the maximal degree is $S^-_{j \rho}$.
Note that the form given in Eq.~\eqref{eq:bio_rate} includes mass-action law as a special case where $D_\rho (\boldsymbol{c}) = 1$.


\subsection{Metzler property of the Jacobian matrix}

Let us assume that the CRN admits at least one fixed point corresponding to the concentration vector $\boldx^\star$. Now, we introduce a perturbation around this fixed point $\boldx^\star$, $\boldx = \boldx^\star + \delta \boldx$, such that the kinetic equations become:
\begin{equation}
\label{eq:linear_dynamics}
\d_t \delta \boldx = \M \cdot \delta \boldx,
\end{equation}
where $\M = \left\lbrace M_{x x^\prime} \right\rbrace$ denotes the Jacobian matrix evaluated at $\boldx^\star$, which can be expressed as
\begin{equation}
\M = \S \cdot \left. \dfrac{\partial \boldsymbol{f}}{\partial \boldx} \right|_{\boldx^\star}.
\end{equation}
This gradient, which contains the response coefficients of the fluxes, is called elasticities (or elastic coefficients) and is a central quantity in the literature on metabolic control analysis (MCA) \cite{Wolf_2010, Heuett2008}.

Let us first show that under mild assumptions about the CRN, the diagonal elements of this Jacobian matrix are negative (see details in appendix \ref{sec:Jacobian}).
We find that this property holds when the CRN satisfies the two following conditions: The first condition is that the chemical network is such that reactants (resp. products) of a given reaction are not simultaneously also products (resp. reactants) of the same reaction, which corresponds to the \textit{non-ambiguity} condition introduced by  Blokhuis \textit{et al.} \cite{Blokhuis2020}.
Mathematically, this is expressed as
\begin{equation}
\label{eq:non_ambiguity}
\forall \rho, ~ S^\pm_{i \rho} > 0 ~ \Rightarrow ~ S^\mp_{i \rho} = 0.
\end{equation}
The second condition is that fluxes depend on concentrations in such a way that they are increasing functions of the reactant concentrations and, conversely, decreasing functions of the product concentrations, reflecting product inhibition. 
In other words, this assumption reads:
\begin{align}
    \label{eq:condition_rate}
    S^+_{i \rho} > 0 \Rightarrow \partial_{c_i} f_\rho \geq 0
    & & \text{and} & &
    S^-_{i \rho} > 0 \Rightarrow \partial_{c_i} f_\rho \leq 0,
\end{align}
a condition that can be checked in particular for  the rate law given by \ref{eq:bio_rate}.
Taken together, these two conditions guarantee the negativity of the diagonal elements of the Jacobian. 

Let us now introduce the central assumption of this paper, namely that the Jacobian matrix evaluated at $\boldx^\star$ should be a Metzler matrix. 
In other words, all the off-diagonal elements in the Jacobian matrix are positive: 
\begin{equation}
    \forall x \neq x^\prime, ~~ M_{x x^\prime} \geq 0. 
\end{equation}
This condition is
satisfied in particular in the special case where the fixed point is the zero concentration point, i.e. in the dilute regime. 
In the absence of an external influx of species into the system, the point of zero concentration, $\boldx^\star = \boldsymbol{0}$, becomes a fixed point of the kinetic equations. 
The stability of the zero fixed point is important on origin of life studies, as it determines whether the system can initiate growth or instead undergo decay.
In the dilute regime, $\boldx = \delta \boldx > 0$ element-wise, making the dynamics of the perturbation a positive linear system; hence, the Jacobian matrix must be Metzler in the dilute regime \cite{farina2011}.

Outside the dilute regime,  the Jacobian matrix is generally non-Metzler, yet there may be certain regimes of parameters (such as the concentrations of species) under which the Jacobian matrix is still Metzler. 
This is fully compatible with the proof mentioned before that there exists a choice of kinetic parameters for which the dynamics of a stoichiometric autocatalytic chemical network can be dominated by the autocatalytic part, when the kinetics is parameter-rich  \cite{stadler2024}.
In the appendix \ref{sec:Jacobian}, we show that the Jacobian matrix is necessarily Metzler if (i) product inhibition is absent; and (ii) if all the reactions involve only one type of species as a reactant (see appendix \ref{sec:Jacobian} for a more precise definition).

\subsection{A Markov chain approach}

According to the Perron–Frobenius theorem for irreducible Metzler matrices \cite{meyer2023}, there exists a unique eigenvalue $\lambda$, which is dominant and which is associated with an eigenvector $\boldsymbol{u}$ whose entries are all strictly positive.
As a result, in the long run, the perturbation tends to align with this leading eigenvector: $\delta \boldx = e^{\lambda \, t} \ \boldsymbol{u} + \boldsymbol{o} \left( e^{\lambda \, t}  \right)$.
In general, deriving an explicit analytic expression for $\lambda$ as a function of the system's parameters (e.g., stoichiometric coefficients, rate constants, chemostatted concentrations) is not feasible.

Since the Jacobian matrix $\M$ shares a similar structure with the generator of continuous-time Markov chains, we can define \textit{transition weights} between species $x'$ and $x$ in a manner similar to transition probabilities in continuous-time Markov chains \cite{norris1998}:
\begin{equation}
	\label{eq:weight_one_step}
	w_{x' \to x} = \dfrac{M_{x x'}}{-M_{x' x'}} ~~\text{for}~~ x\neq x'.
\end{equation}
Because the rows of the Jacobian matrix do not sum to one in our case, the distribution formed by all the transition weights exiting species $x^\prime$ is not a probability distribution: 
\begin{equation}
	\sum_{\substack{x \neq x^\prime} } w_{x^\prime \to x} = K_{x^\prime} \neq 1.  
\end{equation}
Nevertheless, by scaling these weights, we can construct a proper probability distribution:
\begin{equation}
	p_{x^\prime \to x} = \dfrac{w_{x^\prime \to x}}{K_{x^\prime}}. 
\end{equation}
As a result, the dynamics of the linearized network can be related to that of a single particle on a Markov chain whose vertices are the chemical species of the linearized network and whose transition probabilities are given by $p_{x^\prime \to x}$.
In particular, a trajectory $\Gamma= \left\lbrace x_1 \to x_2 \to \cdots \to x_N \right\rbrace$ of the linearized network is associated with the weight 
\begin{equation}
	\mathbb{W} [ \Gamma ] = \prod_{k=1}^{N-1} p_{x_k \to x_{k+1} },
\end{equation}
which is related in the following way to the probability of the same path for the particle in the Markov chain:
\begin{equation}
	\mathbb{P} [ \Gamma ] = \prod_{k=1}^{N-1} p_{x_k \to x_{k+1}} = 
	\prod_{k=1}^{N-1} \hspace{-.1em} \left( K_{x_k} \right)^{-1} \ \mathbb{W} [ \Gamma ]. 
\end{equation}

\subsection{Condition for stability}

For any pair of species $a$, $b$ in the network, we let 
\begin{equation*}
	\mathcal{E} (a \to b) = 
	\big\lbrace 
	a \hspace{-.5mm} \to x_1 \hspace{-.5mm} \to \cdots \to \hspace{-.5mm} x_\ell \hspace{-.5mm} \to \hspace{-.5mm} b 
	\big| \, \ell \geq 0 \hspace{.5mm} ; \hspace{.5mm}
	\forall n, \hspace{.5mm} x_n \neq  b	\big\rbrace, 
\end{equation*}
be the set containing all the paths of arbitrary length which start at $a$ and terminate at $b$, without having visited $b$ at any intermediate step.
Moreover, we introduce the total weight of all these paths:
\begin{equation}
	\label{eq:def_fancy_W}
	\W (a \to b) = \sum_{\Gamma \in \mathcal{E}(a \to b)  } \mathbb{W} [ \Gamma ]. 
\end{equation}
Paths in the set $\mathcal{E}(a \to a)$  are known as  \textit{$a$-based excursions} in the Markov chain literature \cite{norris1998}.
In this context, $\W(a \to a)$ represents the \textit{total excursion weight}. 
Further, the sign of the leading eigenvalue $\lambda$ is directly related to $\W(a \to a)$ through the following criterion, 
\begin{equation}
	\label{eq:criterion}
	\lambda > 0 \hspace{1em}\Leftrightarrow\hspace{1em} \mathcal{W} (a \to a) > 1,
\end{equation}
a result originally derived in \cite{unterberger_stoechiometric_2022}. Due to the importance of this result for this paper, we recall in the appendix \ref{sec:Proof} the main steps of the proof. Note that in addition to the Metzler property, it is also required that the matrix $\M$ corresponds to an irreductible network.

Note that the criterion is independent of the choice of the state $a$ in the network.  
Additionally, as shown in the appendix \ref{sec:first_step} and in the Example 1 below, $\mathcal{W} (a \to a)$ is the solution of a linear problem, as opposed to $\lambda$ which is the root of a polynomial of  arbitrary order.

\subsection{Example 1: a simple autocatalytic cycle in the dilute regime}

To illustrate the above discussion, we consider the following simple autocatalytic chemical network in the dilute regime. 
Specifically, we consider the simplest autocatalytic scheme in which a species $R$ duplicates after two intermediate steps and is subjected to a degradation: 
\begin{align}
	\label{eq:network_example}
	\ce{$R$ <=> $X$
   			<=> $Y$
			<=> $2R$
		}, & \quad &
	\ce{$R$ -> $\varnothing$}. 
\end{align}
Such a network has been used to describe the Breslow autocatalytic cycle in the formose reaction network \cite{Lu2024}. 
The corresponding stoichiometric matrix is then
\begin{equation}
	\S = ~
    \begin{pNiceMatrix}[
		first-col,
		last-row,
		code-for-first-col = \scriptstyle,
		code-for-last-row = \scriptstyle
	]
		R & -1 & 0 & 2 & -1 \\[.3em]
		X & 1 & -1 & 0 & 0 \\[.3em]
		Y & 0 & 1 & -1 & 0 \\[.3em]
		  & 1 & 2 & 3 & 4
	\end{pNiceMatrix},
\end{equation}
where the last column contains the degradation of species $R$. 
If $\boldsymbol{j} = \left(j_1, \, j_2, \, j_3, \, j_4 \right)^\top$ are the net fluxes of the reactions, the kinetic rate equations Eq.~\eqref{eq:kinetic_eqn_genral} yields: 
\begin{align}
	\label{eq:example_dyn}
	\begin{gathered}
		\begin{aligned}
			\d_t r = -j_1 +2 \, j_3 - j_4,
			& \quad & 
			\d_t x = j_1 - j_2 ,
		\end{aligned} \\[.3em]
		\d_t y = j_2 - j_3. 
	\end{gathered}
\end{align}
Now, assuming that each chemical reactions and the degradation proceed with mass-action rate, the fluxes can be expressed as
\begin{equation}
\begin{gathered}
	\begin{aligned}
		j_1 = k_1^+ \, r - k_1^- \, x,
		& \quad & 
		j_2 = k_2^+ \, x - k_2^- \, y,
	\end{aligned} \\[0.5em]
	\begin{aligned}
		j_3 = k_3^+ \, y - k_3^- \, r^2,
		& \quad & 
		j_4 = \kappa \, r, 
	\end{aligned}  
\end{gathered}
\end{equation} 
where $k^+_i$ (resp. $k^-_i$) are the kinetic rate constants of the forward (resp. backward) reactions, and $\tau = \kappa^{-1}$ is the typical timescale associated with the degradation of species $R$. 

Under such assumptions, the zero-concentration point is a fixed point of the kinetic equations Eq.~\eqref{eq:example_dyn}. 
Hence, the sign of $\lambda$ in the dilute regime assesses whether the network in Eq.~\eqref{eq:network_example} is able to trigger exponential growth, for a given set of rate constants and degradation rate. 
In the dilute regime, the Jacobian matrix is
\begin{equation}
	\label{eq:example_lin_dyn}
	\M = ~ 
	\begin{pNiceMatrix}[
		first-col,
		last-row,
		code-for-first-col = \scriptstyle,
		code-for-last-row = \scriptstyle
	]
		R & - k_1^+ - \kappa &          k_1^- 		   & 2 \, k_3^+ 	\\[0.3em]
		X & k_1^+			 & - k_2^+ - k_1^-  &    k_2^-		\\[0.3em]
		Y & 0				 &          k_2^+		   & - k_3^+ - k_2^- \\[.3em]
		  & R & X & Y 
	\end{pNiceMatrix},
\end{equation}
As expected, in the dilute regime, the inhibitory reactions \ce{$2R$ -> $Y$} is absent and the Jacobian matrix is Metzler. 
Further, it is a  non-conservative Markov generator.
Indeed, the first column of $\M$ sums up to $-\kappa$, which is strictly negative, since the degradation contributes to decrease the total amount of species; while the last column of $\M$ sums up to $k_3^+$, which is strictly positive, because the last reaction \ce{$Y$ -> $2R$} increases the total amount of species. 

\begin{figure}[t!]
	\centering
	\vspace*{1em}
	\includegraphics[scale=.7]{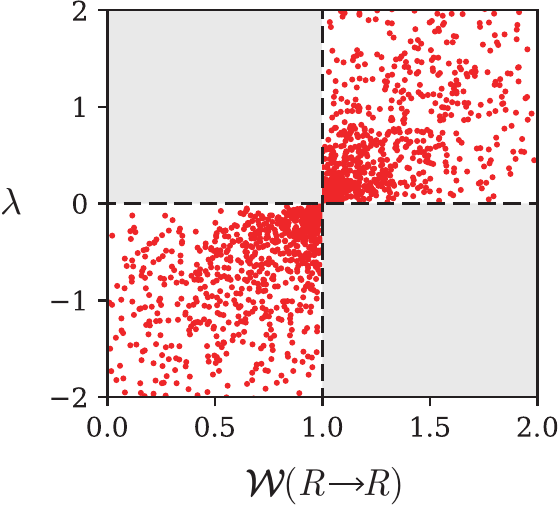}
	\caption{\label{fig:example}
		The criterion for exponential growth Eq.~\eqref{eq:criterion} can be visualized in the $(\W, \, \lambda)$ plane for randomly chosen sets of kinetic rate constants and degradation rate.  
		In the figure, 1000 samples of the kinetic rate constants and degradation rate are represented, each rates was randomly drawn from a log-normal distribution. 
	}
\end{figure}
Applying the definition Eq.~\eqref{eq:weight_one_step}, the elementary transition weights are:
\begin{equation}
	\begin{gathered}
		\begin{aligned}
		w_{R \to X} & = \dfrac{k_1^+}{k_1^+ + \kappa}, & \quad & 
		w_{X \to Y} & = \dfrac{k_2^+}{k_2^+ + k_1^- }, \\[0.3em]
		w_{Y \to R} & = 2 \, \dfrac{k_3^+}{k_3^+ + k_2^- }, & \quad &  
		w_{X \to R} & = \dfrac{k_1^-}{k_2^+ + k_1^- }, 
	\end{aligned} \\[0.3em]
	w_{Y \to X} = \dfrac{k_2^-}{k_3^+ + k_2^- }.
	\end{gathered}
\end{equation} 
Then, the total excursion weight $\W (R \to R)$ is the solution of a linear problem, obtained using a \textit{first step method}, similar to the one use for the computation of the mean first passage time in Markov chains: 
\begin{equation}
	\begin{gathered}
			\W (R \to R) = w_{R \to X} \, \W(X \to R), \\[0.3em]
		\W (X \to R) = w_{X \to R} + w_{X \to Y} \, \W(Y \to R), \\[0.3em]
		\W (Y \to R) = w_{Y \to X} \, \W(X \to R) + w_{Y \to R}.
	\end{gathered}
\end{equation}
Note that the $\W$s in the right hand side of the equations are defined in Eq.~\eqref{eq:def_fancy_W}. 
As a result, the total excursion weight is easily written in terms of the elementary weights and the rate constants:
\begin{align}
	\label{eq:W_example}
	\begin{aligned}
		\W (R \to R) 
		& =
		\dfrac{
			w_{R\to X} \left(  w_{X\to R} + w_{X\to Y} w_{Y\to R} \right) 
		}{
			1 - w_{X \to Y} w_{Y \to X}
		} \\[0.5em]
		& = 
		\dfrac{
			k_1^+ + \kappa_c
		}{
			k_1^+ + \kappa
		},
	\end{aligned}
\end{align}
with $\kappa_c = k_1^+ \, k_2^+ \, k_3^+ / \left( k_1^- \, k_2^- + k_1^- \, k_3^+ + k_2^+ \, k_3^+ \right) $.
Hence, starting with an initially small amount of $R$, the criterion for the system to trigger its exponential growth (\textit{i.e.} $\lambda > 0$) is:
\begin{equation}
	\W (R \to R) > 1 
	~~ \Leftrightarrow ~~
	\kappa < \kappa_c. 
\end{equation}
This relationship is represented in Figure \ref{fig:example} by plotting points on the ($\W$, $\lambda$) plane for randomly generated rate constants. 
Note that, obtaining the above threshold directly from the growth rate $\lambda$ is significantly less convenient because the characteristic polynomial is third-order, 
\begin{equation}
	-\lambda^3 - \alpha_2 \ \lambda^2 - \alpha_1 \ \lambda - k^+_1 k^+_2 k^+_3 \left( 1 - \dfrac{\kappa}{\kappa_c} \right) = 0,
\end{equation}
where $\alpha_2 = k^+_2 + k^+_3 + k^-_1 + k^-_2 $ and $\alpha_1 = k^+_1 k^+_2 k^+_3 /\kappa_c \ + (k^+_2 +k^+_3 + k^-_2) (\kappa + k^+_1) + \kappa k^-_1$. 
For arbitrary rate constants, the solutions given by the cubic formula are unfortunately not tractable. 
However, in this simple case, we can obtain the threshold for the onset of growth Eq.~\eqref{eq:criterion} directly from the characteristic polynomial.
Indeed, by noticing that $\alpha_2$, $\alpha_1 > 0$, if $\kappa < \kappa_c$, all the coefficients in the polynomial are negative and, as a result, the latter cannot admit positive solution implying necessarily that $\lambda < 0$.

The existence of a threshold for the onset of autocatalysis is a general feature that should be expected beyond this specific example. 
Indeed, for any network in our framework $\W (R \to R)$ is a sum of positive monomials of $w_{i \to j}$, but since $w_{i \to j}$ are decreasing functions of the degradation rate $\kappa$, a finite threshold in $\kappa$ must exist below which $\W (R \to R) > 1 $.

\section{Onset of growth in autocatalytic cycles}

In what follows, we apply our criterion to derive general conditions upon which autocatalytic networks initiate growth in the low concentration regime of autocatalytic species (i.e. the dilute regime). 

\subsection{Unicyclic autocatalytic cycles}

\begin{figure*}[t]
	\includegraphics{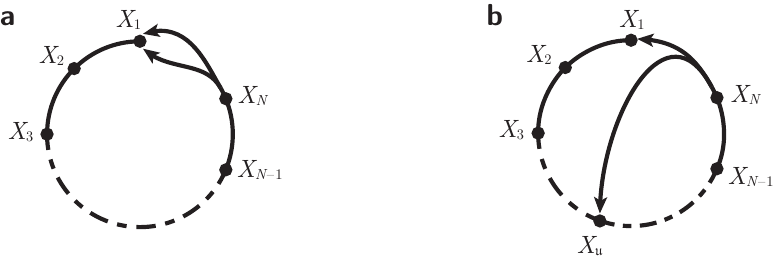}
	\caption{\label{fig:Ind_cycle}
		Uncoupled autocatalytic networks consists of an arbitrary number of intermediates species that reacts reversibly or irreversibly in a linear pathways.  
		\textbf{\textsf{(a)}} In a unicyclic autocatalytic network, the last intermediate, $X_N$, gives back $\mathfrak{p}$ of the first species ; while \textbf{\textsf{(b)}} in a Type~II network, it gives back one unit of $X_1$ and one other unit of an arbitrary upstream species, $X_\mathfrak{u}$. 
	}
\end{figure*}

Let us now focus on the simplest possible topology for an autocatalytic cycle, namely that of a single isolated cycle, composed of $N$ unimolecular reaction steps, involving intermediate species $X_i$ and ending with a single branching point as shown in Figure \ref{fig:Ind_cycle}\textsf{\textbf{a}}. In the last reaction, species $X_N$ splits, producing $\mathfrak{p}$ of the first species $X_1$.
In Ref.~\cite{Blokhuis2020} this topology was classified as the Type~I core. 
We first consider the case of irreversible reactions and then the case of reversible reactions.

\subsubsection{Irreversible case}

In the irreversible case, all the backward rates constants, $k^-_i$, are set to zero and only the forward weights remain, which can be expressed in terms of the specificity  of the reaction of step $i$, namely $s_i=k_i^+/(k_i^++\kappa_i)$ \cite{King1982}. These specificities quantify the probability to move to the next step in the reaction path as opposed to finish there due to side-reaction or death.

With this simple topology, it is straightforward to obtain the total excursion weight:
\begin{equation}
	\label{eq:King_excursion_weight}
	\mathcal{W}  \left( \X_1 \to \X_1 \right) = \mathfrak{p} \prod_{i=1}^N s_i.
\end{equation}
Thus, the criterion Eq.~\eqref{eq:criterion} to achieve exponential growth reads~: 
\begin{equation}
	\label{eq:King_criterion}
	\lambda > 0 
	~~ \Leftrightarrow ~~ 
 	\prod_{i=1}^N s_i  > \dfrac{1}{\mathfrak{p}}.
\end{equation}
This condition was first derived by King \cite{King1982} for the particular case $\mathfrak{p} = 2$, and  
was then further discussed for arbitrary $\mathfrak{p}$ by Szathmary \cite{szathmary_origin_2006}.

Note that the case $\mathfrak{p}=1$  describes for instance a unicycle catalytic network, which is composed of a succession of unimolecular reactions and degradation steps. 
In this case, $\lambda$ is the largest eigenvalue of the Jacobian and it can be shown that the lifetime of the catalyst is then simply $-1/\lambda$ \cite{Hideshi2024}. The practical use of this method has been demonstrated in that work for the case of manganese oxide ($MnO_2$) in experimental reactions in which it is used as a catalyst.

Since $0 < s_i < 1$, the implication of King's criterion is that smaller cycles are more likely 
to grow than bigger ones. The corresponding critical size of these cycles reads, 
\begin{equation}
	N \le -\dfrac{
		\log  \mathfrak{p}
	}{
		\left\langle \log s_i \right\rangle
	},
\end{equation}
where $\left\langle \log s_i \right\rangle = N^{-1} \sum_i \log s_i $ represents the logarithm of the geometric mean of the weights $s_i$ \cite{szathmary_origin_2006}.
 
An equivalent statement to these formula is that the total excursion weight is upper-bounded by the connectivity of the single branching point measured by $\mathfrak{p}$,
\begin{equation}
	\label{eq:bound_W_irreversible}
	\mathcal{W} \left( \X_1 \to \X_1 \right) \leq \mathfrak{p},
\end{equation} 
and the bound is saturated if, and only if, the degradation rates vanish. 
This implies that the total excursion weight of unicyclic and irreversible cycles is intrinsically bounded and cannot reach arbitrary high values 
while no such constraint exists for the growth rate.

\subsubsection{Reversible case}

Now in the reversible case, the backward reactions \ce{$X_i$ -> $X_{i-1}$ } have non-vanishing rates $k^-_i$.
Thus, the forward and backward propensities are  respectively defined as: 
\begin{align}
	\label{eq:propensiities_reversible}
	s^+_i = \dfrac{k_i^+}{k_i^+ + k_i^- + \kappa_i}, 
	& \quad &  
	s^-_i = \dfrac{k_i^-}{k_i^+ + k_i^- + \kappa_i}.
\end{align}
Note that, because there is no backward reaction having $X_1$ as reactant, $k_1^- = 0$.
As before, the elementary weights are related to the propensities~: 
\begin{equation}
	\label{eq:elem_weights_linear}
	\begin{gathered}
	\forall 1 \leq i \leq N-1 ~, ~	w_{X_i \to X_{i \pm 1}}  =  s_i^\pm \ ; \\[.5em]
	w_{X_N \to X_1} = \mathfrak{p} s_N^+.
	\end{gathered}
\end{equation}
Performing a similar method to one presented in the Example 1 above (see the Section 3 of the supplemental materials for details), one can derive the analytic expression of the  total excursion weights, $\W (\X_1 \to \X_1)$:
\begin{equation}
	\label{eq:weight_King_reversible}
	\W  \left( \X_1 \to \X_1 \right) = \dfrac{
		s_1^+ \, s_2^- \, \Delta_3 + \mathfrak{p} \prod_{i = 1}^N s_i^+
	}{
		\Delta_2
	},
\end{equation}
where $\Delta_2$ and $\Delta_3$ are two determinants whose expressions are given in appendix \ref{sec:proof_single_cycle}. 
Our previous result Eq.~\eqref{eq:bound_W_irreversible} extends to the case where backwards reactions are included because we show in this appendix that ~: 
\begin{align}
	\W \left( \X_1 \to \X_1 \right) \leq \mathfrak{p}.
\end{align}
As before, this means that, in the presence of backward reactions, 
the total excursion weight of a single cycle is still limited by the connectivity of the branching point as in the case with no backward reactions. 

Furthermore, assuming that the forward specificities take the same values in the reversible and irreversible version of the same network, the presence of backward reactions has a positive impact on the total excursion weight.
Indeed, the total excursion weight in the irreversible case given by Eq.~\eqref{eq:King_excursion_weight} is always lower than its counterpart with reversible reactions given by Eq.~\eqref{eq:weight_King_reversible}, regardless of the values taken by backward rates as long as the $(s_i^+)$ take the same values in both networks. 
The intuitive explanation is that the total weight contains more paths in the reversible case.
Note that fixing the values of the forward specificities to be the same in both networks imply different kinetic rates constants in both networks. 

If we now fix instead the values of the forward rate constants, we recover the irreversible case where, for each reactions, $k_i^+ \gg k_i^-$~:
\begin{equation}
	\label{eq:behavior_W_1}
	\W (X_1 \to X_1) \longrightarrow \mathfrak{p} \prod_{i=1}^N \dfrac{k_i^+}{k_i^+ + \kappa_i}.
\end{equation}
In the regime where the backward rates dominate, ${k^+_i \ll k^-_i}$, the total excursion weight is given by the elementary weight of the first reaction~:
\begin{equation}
	\label{eq:behavior_W_2}
	\W (X_1 \to X_1)  \longrightarrow s_1^+ = \dfrac{k_1^+}{k_1^+ + \kappa_1} \leq 1.
\end{equation}
Not surprisingly, in this regime the network goes extinct (i.e. $\lambda <0$) or, if there is no degradation on the first species, it neither grows nor decreases (i.e. $\lambda = 0$ and $\W = 1$).

\subsection{Unicyclic network with side-branches}

When a side branch is introduced in an unicyclic network with $\mathfrak{p} = 1$, one obtains a Type~II network in the classification of Ref.~\cite{Blokhuis2020}.
Such network is represented in Fig.~\ref{fig:Ind_cycle}\textsf{\textbf{b}}: one can see that there is a side-branch that connects species $X_N$ back to species $X_\mathfrak{u}$. 
In practice, this class of networks is large because there is a lot of freedom regarding the content (in terms of reactions) and length of this side branch. 

Let us consider again first the case of irreversible reactions and then the case of reversible reactions in an example.

\subsubsection{Irreversible case}
By the same method used in previous sections, one can obtain an explicit form of the total excursion weight in the irreversible case:
\begin{equation}
	\label{eq:typeII}
	\W (X_1 \to X_1) = \dfrac{
		\prod_{i=1}^N s_i^+
	}{
		1 - \prod_{i=\mathfrak{u}}^N s_i^+
	}.  
\end{equation}
Some observations are, in order : 
(i) the side branch plays an essential role in these Type~II networks, without it, the total weight would be $\prod_{i=1}^N s_i^+$ and thus would remain below threshold since $0 < s_i^+ < 1$;
(ii) Type~II networks are intrinsically more robust than Type~I networks, because the total excursion weight of Type~I networks is limited by the vertex connectivity $\mathfrak{p}$, which in practice can not be a very large number, while the total excursion weight of Type~II networks is enhanced by the contribution of the side branch. 
More precisely, if we assume that all $s_i^+$ take the same values in both networks, the total excursion weight of Type~II networks overcomes that of Type~I when 
\begin{equation}
	\prod_{i=\mathfrak{u}}^N s_i^+ > 1 - \frac{1}{\mathfrak{p}}.
\end{equation}

Naturally, the benefice of side branches can only go so far, and the total weight of Type~II networks still remains finite. Before, we present another strategy for further enhancing network connectivity and robustness thanks to coupling autocatalytic cycles to each other, let us study a particular example of a simple network with a side-branch. 

\subsubsection{Reversible case}

Let us discuss the case of reversible reactions on a second example.
The simplest autocatalytic network containing a side-branch is a type II autocatalytic network in the classification of Ref. \cite{Blokhuis2020}:
\begin{align}
	\label{eq:network_example2}
	\ce{$A$ <=> $B$}, & \quad &
	\ce{$B$ <=> $C$}, & \quad & 
	\ce{$C$ <=> $A$ + $B$}.
\end{align}
For the kinetics, let us assume a general rate law not necessarily of the form of mass-action law kinetics. 
To do that, we consider that unidirectional fluxes are described by arbitrary function $f_{\pm i} ( \cdot ) $ which depend solely on the reactant of the reactions but are increasing functions of concentrations.  
Note that these assumptions are compatible with a large panel of experimentally relevant rate laws as reviewed in Ref.~\cite{Wolf_2010}. 
Hence, the net fluxes of the reactions can be written
\begin{equation}
\begin{gathered}
		\begin{aligned}
		j_1 = f_{+1} (a) - f_{-1} (b), 
		& \quad & 
		j_2 = f_{+2} (b) - f_{-2} (c),
	\end{aligned} \\[.5em]
	j_3 = f_{+3} (c) - f_{-3} (a, \ b), 
\end{gathered}
\end{equation} 
where $a$, $b$ and $c$ are the concentration of species $A$, $B$ and $C$, respectively.
From this, the Jacobian matrix at an arbitrary point is
\begin{widetext}
\begin{equation}
	\M = ~
\begin{pNiceMatrix}
	- f'_{+1}(a) - \partial_a f_{-3}(a,b) & f'_{-1}(b) - \partial_b f_{-3}(a,b) & f'_{+3}(c) \\[.8em]
	f'_{+1}(a) - \partial_a f_{-3}(a,b) & -f'_{-1}(b) - f'_{+2}(b) - \partial_b f_{-3}(a,b) & f'_{-2}(c) + f'_3(c) \\[.8em]
	\partial_a f_{-3}(a,b) & f'_{+2}(b) + \partial_b f_{-3}(a,b) & -f'_{-2}(c) - f'_{+3}(c)
\end{pNiceMatrix}.
\end{equation}
\end{widetext}
Hence, assuming the unidirectional fluxes are increasing function of the concentration, the Jacobian matrix is Metzler if, and only if, 
\begin{equation}
\label{eq:Metzler-conditions}
	f'_{+1}(a) > \partial_a f_{-3}(a,b)
	\hspace{1em} \text{and} \hspace{1em}
	f'_{-1}(b) >  \partial_b f_{-3}(a,b).
\end{equation}

In other words, the Jacobian matrix is Metzler when the response coefficient of the inhibitory reaction \ce{$A$ + $B$ ->  $C$} with respect to a change of $a$ is lower than the ones of the reactions \ce{$A$ ->  $B$} and \ce{$B$ -> $A$}. 

For instance, when mass-action law is used, these conditions translate into lower bounds that need to be satisfied by $a$ and $b$. For this case, we can write $f_{\pm i}(x)=k_i^\pm x$ and $f_{\pm i}(x,y)= k_i^\pm  x y$ where $k_i^\pm$ are rate constants and $x$ and $y$ are concentrations. 
Then, the conditions of Eq.~\eqref{eq:Metzler-conditions} corresponds to the regime of concentrations such that $b < k_1^+/k_{3}^-$ and $a< k_{1}^- /k_{3}^-$. 

If we consider again the case of irreversible reactions, the condition 
of Eq. \ref{eq:Metzler-conditions} necessarily hold since $f_{-3}$ vanish. The response coefficients, namely the functions $f'_i$ then play an equivalent role as the forward rate constants in the expression of the specificies $s_i$ given in the previous section.

\subsection{Benefit of network coupling}

\begin{figure*}[t!]
	\includegraphics{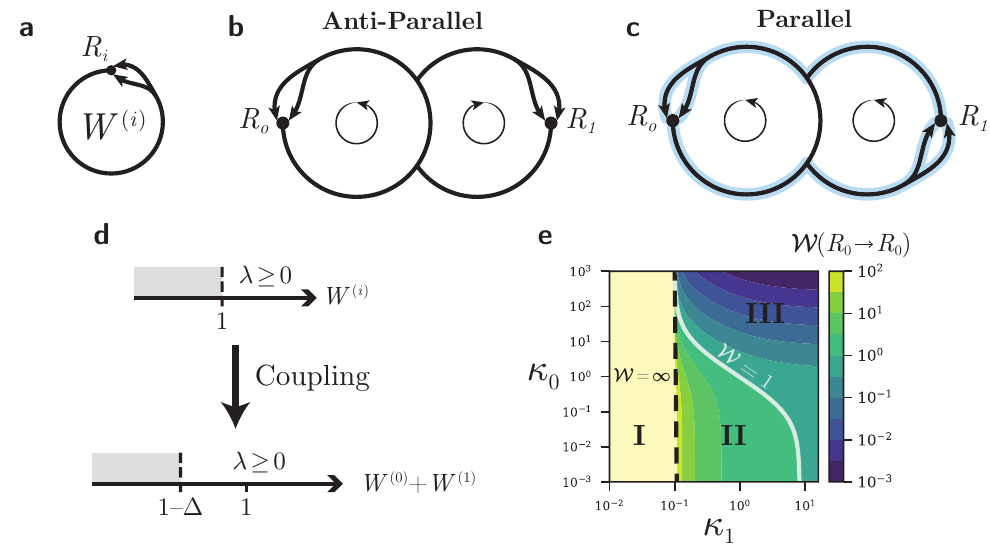}
	\caption{\label{fig:coupling}
		Cycle coupling facilitates the onset of growth. 
		\textbf{\textsf{(a)}} 
        Without coupling, the unicyclic cycle associated to $R_i$ has a total excursion weight of $W^{(i)}$. 
		The coupling can be done in two different ways:
		\textbf{\textsf{(b)}} in the anti-parallel coupling, the two cycles are running in opposite direction, this adds the possibility to loop an arbitrary number of time in cycle $(1 -i)$ before going back to $R_i$; 
		\textbf{\textsf{(c)}} in the parallel coupling the two cycles are running in the same direction, this adds a new class of excursion (highlighted in blue) that exists only in the coupled network. 	
		\textbf{\textsf{(d)}} For the independent unicyclic cycle the threshold is $W^{(i)} \geq 1$ while, after the coupling, the $W^{(i)}$s of the two cycles sum up. 
		In addition, for parallel coupling, due to the presence of a new excursion class, the threshold is affected by $\Delta$ (here represented for $\Delta > 0$). 
		In the anti-parallel case, $\Delta = 0$.
		\textbf{\textsf{(e)}} When $\kappa_1$ is below a certain threshold (black dotted line), $W^{(1)} > 1$, hence, cycle $(1)$ supports the growth of the coupled cycle for any value of $W^{(0)}$ and $\mathcal{W} (R_0 \to R_0)$ diverges. 
		While, above this threshold, there exists a region were the two cycles cooperate ($W^{(0)} < 1$ and $W^{(1)} < 1$) to achieve exponential growth, this region lies between the black dotted curve and the solid white curve.
		Plot \textbf{\textsf{(e)}} was generated with: 
		$k_1^+ = \nu_1^+ = k_1^- = \nu_1^- = k_2^- = 1$, $k_3^+ = 8$, $\nu_3^+ = 10$ and $k_2^+ = 100$ such that, when $\kappa_0 = \kappa_1 = 0$, one has $W^{(0)} \approx 0.88 < 1$ and $W^{(1)} \approx 1.09 > 1$.
	}
\end{figure*}

This idea of coupling autocatalytic cycles to each other has been considered early in Origin of Life studies; this is essentially the idea of hypercycles introduced by Eigen in the seventies \cite{eigen2012hypercycle}.
To illustrate that case, we consider two unicyclic autocatalytic networks similar to Eq.~\eqref{eq:network_example} with two intermediate species $X$ and $Y$ shared between the two cycles. 
The last reaction being irreversible in the linear regime, it imposes a direction in both networks. 
This implies that, coupling can occur in two different ways: 
(i) the \textit{anti-parallel coupling} occurs when the cycles runs in opposite direction once coupled (see Fig.~\ref{fig:coupling}\textsf{\textbf{b}}),
\begin{equation}
	\label{eq:parallel}
	\begin{gathered}
		\ce{
		$R_0$ <=>[$k^+_1$][$k^-_1$] $X$
		<=>[$k^+_2$][$k^-_2$] $Y$
		->[$k^+_3$] 2$R_0$
	}, \\
	\ce{
		$R_1$ <=>[$\nu^+_1$][$\nu^-_1$] $X$
		<=>[$k^+_2$][$k^-_2$] $Y$
		->[$\nu^+_3$] 2$R_1$
	} ~~ ;
	\end{gathered}
\end{equation}
(ii) the \textit{parallel coupling} arise when both are running in the same direction (see Fig.~\ref{fig:coupling}\textsf{\textbf{c}}), 
\begin{equation}
	\label{eq:anti-parallel}
	\begin{gathered}
		\ce{
		$R_0$ <=>[$k^+_1$][$k^-_1$] $X$
		<=>[$k^+_2$][$k^-_2$] $Y$
		->[$k^+_3$] 2$R_0$
	}, \\
	\ce{
		$R_1$ <=>[$\nu^+_1$][$\nu^-_1$] $Y$
		<=>[$k^-_2$][$k^+_2$] $X$
		->[$\nu^+_3$] 2$R_1$ 
	}.
	\end{gathered}
\end{equation}
Importantly, we will assume equal transition weights for the two independent cycles and the coupled network. 
Hence when the cycles will be considered independently, the reaction starting from the intermediates and going to the other network  is counted as a degradation. 
For example, considering the cycle associated to $R_0$ independently of the one of $R_1$ we should add the following degradation pathways in the anti-parallel case: 
\begin{align}
	\ce{$X$ ->[$\nu_1^-$] $\varnothing$},
	& \quad &
	\ce{$Y$ ->[$\nu_3^+$] $\varnothing$} ; 
\end{align}
while in the parallel case it should be 
\begin{align}
	\ce{$X$ ->[$\nu_3^+$] $\varnothing$},
	& \quad &
	\ce{$Y$ ->[$\nu_1^-$] $\varnothing$} ;
\end{align}
and similarly for cycle (1), replacing the "$\nu$" by "$k$".
When the two cycles are coupled, these degradation become the reactions coupling the cycles together. 
Finally, in all cases, we consider that species $R_i$ is subjected to degradation with rate $\kappa_i$: 
\begin{align}
	\ce{$R_0$ ->[$\kappa_0$] $\varnothing$},
	& \quad &
	\ce{$R_1$ ->[$\kappa_1$] $\varnothing$}.
\end{align} 
Denoting $W^{(i)}$ the total excursion weight of species $R_i$ when its cycle is considered independently from the one of $R_{1-i}$ (see Fig.~\ref{fig:coupling}\textsf{\textbf{a}}), its expression in term of the elementary weights can be derived applying Eq.~\eqref{eq:weight_King_reversible}.
To see the influence of coupling we consider now, $\mathcal{W} (R_i \to R_i)$, the total excursion weight of species $R_i$ once the two cycles are coupled. 
The latter can be easily obtained as a function of the elementary weights using a first-step method and its expression depends on the coupling (anti-parallel/parallel).

\subsubsection{Anti-parallel coupling}

In the anti-parallel coupling, the two classes of excursion of $R_i$ are identical to those of the uncoupled case, $R_i \to X \to Y \to 2 \, R_i$ or $R_i \to X \to R_i$, however, the coupling adds the possibility to loop an arbitrary number of time in cycle $(1-i)$ before reaching $R_i$. 
This is reflected in the expression of the total excursion weight in this case which reads:
\begin{equation}
	\label{eq:anti-parallel_coupling}
	\W (R_i \to R_i) = \dfrac{W^{(i)}}{1 - W^{(1-i)}}.
\end{equation}
In Eq.~\eqref{eq:anti-parallel_coupling}, the numerator accounts for the fact no new class of excursion are present in the coupled network, compared to the uncoupled case; while the denominator describes the possibility of doing an arbitrary number of loops in cycle $(1-i)$.
Doing $n$ such loops brings a weight $(W^{(1-i)})^n$, hence, the total contribution of all the back-and-forth in cycle $(1-i$) is the a geometric sum of $(W^{(1-i)})^n$.
From the expression of the total excursion weight, we can quantify the positive influence of coupling: 
\begin{enumerate}
	\item if $W^{(1-i)} > 1$ or, equivalently, if the cycle of $R_{1-i}$ already ensures exponential growth, the total excursion weight Eq.~\eqref{eq:anti-parallel_coupling} becomes formally negative, meaning that, in truth, it diverges. 
	Indeed, in this case, the geometric summation of $W^{(1-i)}$ diverges.
	This occurs in the region before the black dotted curve in Fig.~\ref{fig:coupling}\textsf{\textbf{e}}.
\item If $W^{(1-i)} < 1$, the coupling lowers the threshold to achieve exponential growth in the coupled network for $W^{(i)}$: 
	\begin{equation}
		\label{eq:thresh_anti-parallel}
		W^{(i)} > 1 - W^{(1-i)}, 
	\end{equation}
	which can be fulfilled for $W^{(i)} < 1$.
	In Fig.~\ref{fig:coupling}\textsf{\textbf{e}}, this case takes place between the black dotted curve and the solid white curve (which is the threshold $\W = 1$).
\end{enumerate}

In the first case, the coupled cycle achieves exponential growth through just one of the two cycles, which \textit{sustains} the overall growth.
In the second case, both cycles, taken independently, are not able to trigger exponential growth but, once coupled, they can \textit{cooperate} to grow as a whole.  
Note that Eq.~\eqref{eq:thresh_anti-parallel} implies that, as far as the threshold is concerned, the excursion weights $W^{(i)}$ behaves in an additive way (cf. Fig.~\ref{fig:coupling}\textsf{\textbf{d}}).

\subsubsection{Parallel coupling}

Unlike the anti-parallel case, parallel coupling adds a new class excursion, $R_0 \to X \to 2 \, R_1 \to Y \to 2 \, R_0$ (and similarly for $R_1$), that exists only in the coupled network. 
The latter is represented as the blue path  in Fig.~\ref{fig:coupling}\textsf{\textbf{c}}. 
In addition, as before, if one looks for ${\W (R_i \to R_i)}$,
parallel coupling also brings the possibility of doing an arbitrary number of loops in network $(1-i)$. 
Consequently, in the parallel coupling the  total excursion weight is:
\begin{equation}
	\label{eq:parallel_coupling}
	\mathcal{W}(R_i \to R_i) = \dfrac{W^{(i)}}{1 - W^{(1-i)}} + \dfrac{\Delta}{ 1 - W^{(1-i)} }, 
\end{equation}
where
\begin{widetext}
	\begin{equation}
	\label{eq:Delta}
	\Delta =  \left. \left. \dfrac{w_{R_0 \to X} w_{R_1 \to Y}}{1  - w_{X \to Y} w_{Y \to X} }
	\right[ w_{Y \to R_0}  w_{X \to R_1} -  w_{X \to R_0} w_{Y \to R_1} \right].
	\end{equation} 
\end{widetext}
In Eq.~\eqref{eq:parallel_coupling}, the first term has the same interpretation as before and the additional term takes into account the new excursion class brought by the coupling. 
Surprisingly, the influence of the new class ($R_0 \to X \to 2 \, R_1 \to Y \to 2 \, R_0$) on the total excursion weight has to be compared to the two classes associated with the backward reactions, $R_0 \to X \to R_0$ and $R_1 \to Y \to R_1$, such that: 
\begin{equation}
	\label{eq:criterion_parallel_coupling}
	\Delta > 0  ~~ \Leftrightarrow ~~  w_{Y \to R_0}  w_{X \to R_1} > w_{X \to R_0} w_{Y \to R_1}.
\end{equation}
This implies that $\Delta > 0$ if the former dominates and, in the converse case, $\Delta < 0$. 
More precisely, writing explicitly the expression of the elementary weights, the condition Eq.~\eqref{eq:criterion_parallel_coupling} can be translated on a condition on the kinetic rate constants: $4 \, k_3^+ \nu_3^+ > k_1^- \nu_1^-$.  

Again, if one has $W^{(i)}<1$ and $W^{( 1- i) } > 1$, cycle $(1 - i)$ support the growth of the coupled cycle, making $\W (R_i \to R_i) >1$ for any value of $W^{(i)}$.
However, in the parallel coupling, if $W^{(1-i)} < 1$ the effective threshold for $W^{(i)}$ to reach exponential growth in the coupled network is altered compared to Eq.~\eqref{eq:thresh_anti-parallel}: 
\begin{equation}
	W^{(i)}  > 1 - W^{(1-i)} - \Delta.
\end{equation}
Hence, positive values of $\Delta$ are even more favorable for the threshold. 
This is not surprising, as $\Delta > 0$ favors the new excursion class compared to the two excursions associated with the backward reactions and the former is the only one that increases the number of species (while the two others are not creating additional species). 

As in the example, the instability threshold of two coupled unicyclic cycles is straightforward using the approach based on the total excursion weight, even though the underlying growth rate, $\lambda$, for the coupled network cannot be obtained analytically (in general, it will be the maximum root of an order 4 polynomial). 
The present approach also makes clear that coupling two autocatalytic cycles to each other can lead to robust networks, potentially more robust than type II networks, because the total excursion weight can diverge when $W^{(1-i)}$ approaches 1, while the total excursion weight of type II networks remains finite.

\subsection{General case} 

Although Eqs.~\eqref{eq:anti-parallel_coupling} and \eqref{eq:parallel_coupling} do not hold for couplings involving more complex networks, they nevertheless capture the essential mechanisms by which coupling promotes the onset of growth.
First, coupling tends to reduce the threshold for growth associated with each of the autocatalytic cycles in the coupled network, enabling cycle cooperation.  
Specifically, if $W^{(0)} < 1$ and $W^{(1)} < 1$ are the total excursion weights of two decaying autocatalytic cycles, the total excursion weight $\W$, obtained by coupling these two cycles, might still be able to overcome the threshold: $\W > 1$. 
Additionally, the growth of one individual cycle in a coupled network is able to sustain the growth of the whole network.
In that case, one of the individual weight already exceeds the threshold, $W^{(i)} > 1$,  such that $\W > 1$ for any value of $W^{(1-i)}$.
Finally, as illustrated in the anti-parallel case above, coupling also facilitates the onset of growth by enabling the emergence of new excursion paths that exist only in the coupled network.

\section{Summary and discussion}

We have presented a framework to calculate the decay threshold of a CRN, above which replication can overcome degradation reactions.
The formalism is solely based on the assumption of a Metzler Jacobian and applies to various rate laws. 
The Metzler property holds in particular for autocatalytic cores, which represent a part of autocatalytic networks that can
be effectively described by linear kinetics. 
This method circumvents the need to compute the eigenvalue spectrum of the Jacobian, which is generally intractable. Instead, it exploits the hypergraph description of a chemical network to extract the cumulative weight of all the excursion paths. 
The latter is shown to be related to the sign of the leading eigenvalue and is computed as a solution of a linear system.
As a result, our criterion can be expressed in a straightforward way even when the parameters associated with the rates are not known. 

Besides helping to determine whether a given network is growing (or is dynamically unstable), the total excursion weight $\W$ can also inform on how far a given network is from the threshold of exponential growth (or instability): high values of $\W$ imply that a perturbation in the network is less likely to disturb the regime of exponential growth, while a value of $\W$ close to threshold may indicate a fragile or sensitive network.

Thanks to this insight, we can formulate a number of design principles for autocatalytic networks, which could have a  potential relevance for Origin of Life studies. First, we find that unicyclic autocatalytic networks are rather fragile and their size is also limited: large cycles should grow more slowly and are thus more likely to be overcome by small cycles in a competitive setting \cite{Konnyu2024}. For a fixed size, perturbations of kinetic parameters in type I networks are expected to easily produce extinction of the cycle as compared with other topologies. By introducing additional pathways in the unicyclic network, either in the form of side branches and/or thanks to coupling between cycles, the robustness can be improved \cite{Blokhuis2020}. Both effects tend to lower the threshold for exponential growth and favor highly connected and intricate networks, as observed in metabolic networks. These two options for improvement with respect to type I cycles are however, not equivalent.

Let us first consider the benefit of side branches. A peculiarity of type I cycles is the presence of a branching point in the network, where a dimer splits into exactly two identical molecules. Typical networks with organic molecules do not often contain such a reaction, which might be the reason why type I cycles are more commonly found in redox chemical networks based on non-organic molecules \cite{Peng2023}. In this paper, we have shown that type II cycles are more robust than type I cycles already at the level of a Markov chain description, which suggests that a transition from a type I cycle to a type II might have occurred in the course of evolution. This could correspond to a transition from a redox chemistry based on non-organic molecules to a non-redox chemistry based on organic molecules.

In this context, it is interesting to note that the reverse TCA cycle is an autocatalytic cycle only because of its side branch \cite{orgel_2008}, which makes it a type II cycle. When exploring all possible cycles operating with the same chemical compounds as the r-TCA cycle, one finds that a significant fraction of them also possess at least one side branch \cite{zubarev_uncertainty_2015}. Further, among known natural or synthetic carbon-fixing pathways, a majority of them possess at least one side branch \cite{Bar-Even_2010}. These observations tend to confirm the view that side-branches are a feature of CRN that is favored by evolution.

Another option to improve robustness is to couple two type I cycles together. Depending on how this coupling is done, various ecological interactions can arise between the two cycles, which can, for instance, be of the competition, mutualism, predation or bistability types \cite{Peng2023, Konnyu2024}. 
Bistability or in general multistability could be an important feature in Origin of life models, because a diversity of stable states allows a form of heritable transmission and chemical selection to take place.

While our formalism can not describe bistability directly, it shows that, irrespective of the details of the coupling and again at a Markov chain level, coupling two cycles together is an even more efficient strategy to improve robustness than the incorporation of side branches because this can lead to larger values (in fact, potentially infinite) of the total excursion weight.
Indeed, we found that a growing autocatalytic cycle can support the growth of the ensemble of the two coupled cycles even when the second cycle taken in isolation would not achieve exponential growth: in this situation, the coupled network achieves growth for arbitrary values of the kinetic rates in the second cycle. This leads to a divergence of the total excursion weight, which can not be achieved with type II topology. 
Further, the possible synergy in the association of the two cycles is reminiscent of free energy transduction in biology and physics, where a thermodynamically unfavorable process can be realized in practice by coupling it with a favorable one. 
This analogy suggests that important thermodynamic concepts, such as that of free energy transduction, could be useful to understand ecosystem dynamics whether at a molecular or at a larger scale \cite{Goyal2023} with important consequences for research on the Origin of Life.

\section*{Acknowledgments}

We acknowledge a critical reading of the paper by N. Vassena and insightful discussions with W. Liebermeister. We thank Wilhelm T.S. Huck for an invitation at Nimwegen University where some key ideas for this work had the opportunity to mature.

\onecolumngrid 

\section{Appendices}

\subsection{\label{sec:Jacobian}Structure of the Jacobian matrix}

Let us first prove that the diagonal elements of the Jacobian are negative under conditions discussed in the main text. Since both stoichiometric coefficients 
$S^\pm_{i \rho}$ are non-negative, one can separate the case where they are zero from the case where they are strictly positive in the following way: 
\begin{equation}
\begin{aligned}
	M_{i i} & = \sum_\rho (S^-_{i \rho} - S^+_{i \rho}) \ \partial_{c_i} f_\rho \\[.5em]
	& =
	\sum_{\substack{\rho \ \text{s.t.} \\ S^-_{i \rho} > 0}} S^-_{i \rho} \ \underbrace{\partial_{c_i} f_\rho}_{\leq 0} 
	- 
	\sum_{\substack{\rho \ \text{s.t.} \\ S^+_{i \rho} > 0}} S^+_{i \rho} \ \underbrace{\partial_{c_i} f_\rho}_{\geq 0} ~~
	< 0.
\end{aligned}
\end{equation}
This splitting of the sum is only possible because the reactions are non-ambiguous, and the conclusion on the sign of the global expression uses the condition Eq.~\eqref{eq:condition_rate} on the rates.
Note that, by writing a strict inequality for the diagonal elements we are ignoring potential absorbing states of the linearized network. 
This amounts to considering only species that are reactant of at least one reaction. 
The off-diagonal coefficients are more complicated to express in the general case: 
\begin{equation}
	\begin{aligned}
		\label{eq:off_digaonal_M}
		M_{i j} & = \sum_\rho (S^-_{i \rho} - S^+_{i \rho}) \ \partial_{c_j} f_\rho = 
		\sum_{\substack{\rho \ \text{s.t.} \\ S^-_{i \rho} > 0}} S^-_{i \rho} \ \partial_{c_j} f_\rho
		- 
		\sum_{\substack{\rho \ \text{s.t.} \\ S^+_{i \rho} > 0}} S^+_{i \rho} \ \partial_{c_j} f_\rho \\
		& = \Bigg(
		\sum_{\substack{\rho \ \text{s.t.} \\ S^-_{i \rho} > 0 , \ S^+_{j \rho} > 0}} \hspace{-1em} S^-_{i \rho} \ \partial_{c_j} f_\rho 
		~ - \hspace{-1em}
		\sum_{\substack{\rho \ \text{s.t.} \\ S^+_{i \rho} > 0, \ S^-_{j \rho} > 0}} \hspace{-1em} S^+_{i \rho} \ \partial_{c_j} f_\rho
		\Bigg) 
		+ 
		\Bigg(
		\sum_{\substack{\rho \ \text{s.t.} \\ S^-_{i \rho} > 0, \ S^-_{j \rho} > 0}} \hspace{-1em} S^-_{i \rho} \ \partial_{c_j} f_\rho 
		~ - \hspace{-1em}
		\sum_{\substack{\rho \ \text{s.t.} \\ S^+_{i \rho} > 0, \ S^+_{j \rho} > 0}} \hspace{-1em} S^+_{i \rho} \ \partial_{c_j} f_\rho
		\Bigg). 
	\end{aligned}
\end{equation}
In the last equality, the first parenthesis contains positive terms only, while the second parenthesis contains negative terms. 

Now, the off-diagonal elements of the Jacobian matrix take a simple form under the limiting case where the reaction rates are not influenced by the products concentrations (i.e. product inhibition is absent): 
\begin{equation*}
    \forall \rho, ~ S^-_{j \rho} > 0 ~ \Rightarrow ~ \partial_{c_j} f_\rho = 0. 
\end{equation*}
In that case, the second and third sum in Eq.~\eqref{eq:off_digaonal_M} vanish: 
\begin{equation}
    \label{eq:limiting_case_I}
    M_{ij} = \hspace{-1em}
	\sum_{\substack{\rho \ \text{s.t.} \\ S^-_{i \rho} > 0 , \ S^+_{j \rho} > 0}} \hspace{-1em} S^-_{i \rho} \ \underbrace{\partial_{c_j}f_\rho}_{\geq 0} 
	\hspace{1em}  - \hspace{-1em} 
	\sum_{\substack{\rho \ \text{s.t.} \\ S^+_{i \rho} > 0, \ S^+_{j \rho} > 0}} \hspace{-1em} S^+_{i \rho} \ \underbrace{\partial_{c_j}f_\rho}_{\geq 0} . 
\end{equation}
Hence, when product inhibition is absent (for example if mass-action kinetics is assumed), the off-diagonal entry $M_{i j}$ is positive if, and only if: 
\begin{equation}
    \label{eq:condition_Metzler}
    \sum_{\substack{\rho \ \text{s.t.} \\ S^-_{i \rho} > 0 , \ S^+_{j \rho} > 0}} \hspace{-1em} S^-_{i \rho} \ \partial_{c_j} f_\rho
    ~ \geq \hspace{-1em}
    \sum_{\substack{\rho \ \text{s.t.} \\ S^+_{i \rho} > 0, \ S^+_{j \rho} > 0}} \hspace{-1em} S^+_{i \rho} \ \partial_{c_j} f_\rho . 
\end{equation}
In other words, $M_{ij} \geq 0$ if the total response of the reactions producing species $x_i$ dominates that of the reactions in which species $x_i$ partakes as reactant along with other species, namely the reactions
\begin{equation}
    \label{eq:reaction_to_remove}
    S^+_{i \rho} \ x_i + S^+_{j \rho} \ x_j + \cdots \ce{->} \cdots  
\end{equation}
with $S^+_{i \rho} > 0$ and $S^+_{j \rho} > 0$. 
As exemplified in Example 2 above, Eq.~\eqref{eq:condition_Metzler} can be written as a condition on the parameters of the rates and on the concentrations. 

Note that, if the network is deprived of reactions in form of Eq.~\eqref{eq:reaction_to_remove} then
\begin{equation}
     M_{ij} ~  = \hspace{-.5em}
	\sum_{\substack{\rho \ \text{s.t.} \\ S^-_{i \rho} > 0 , \ S^+_{j \rho} > 0}} \hspace{-1em} S^-_{i \rho} \ \partial_{c_j} f_\rho ~  \geq 0. 
\end{equation}
Then, if the set containing all the reactions in the form of  Eq.~\eqref{eq:reaction_to_remove} is empty (i.e. 
$\left\lbrace \rho ~ \text{s.t.} ~ \exists i \neq j;~  S^+_{i \rho} > 0 \, ~{\rm and}~ S^+_{j \rho} > 0 \right\rbrace = \varnothing$), all reactions in the network can be written as
$$ S^+_{i \rho} \ x_i \ce{->} \cdots, $$
and all the off-diagonal entries in $\M$ are positive, resulting in a Metzler Jacobian matrix. 
For example, the reversible unicyclic (type I) networks, 
$ X_1 \ce{<-->} \cdots \ce{<-->} X_n \ce{<-->} \mathfrak{p} X_1,
$
fall in this class.

\subsection{\label{sec:Proof}Proof of the criterion}

The formal solution of Eq.~(3) in the main text is
\begin{equation}
	\boldx (t) = e^{\M \, t} \cdot \boldx_0.
\end{equation}
Taking the Laplace transform of the solution provides, 
\begin{equation} 
	\hat{\boldx} (s) = 
	\left( 
	\int_0^{+ \infty} \hspace{-1.5em} \d t ~
	e^{t \left(\mathbb{M} - s \mathbbm{1} \right) } 
	\right) \cdot 
	\boldx_0.
\end{equation}
When it converges, the integral in the RHS matches the resolvent of $\M$:
\begin{equation}
	\label{eq:def_resolvent}
	R (s) 
	= \int_0^{+ \infty} \hspace{-1.5em} \d t ~ e^{t \left( \mathbb{M} - s \mathbbm{1} \right) } 
	= \left(s \mathbbm{1} - \mathbb{M} \right)^{-1} 
\end{equation}
The latter is tightly related to the spectrum of $\M$ as, whenever $s \in \mathrm{Sp}\left[ \M \right]$, the resolvent has a singularity. 
Furthermore, the entries of the Laplace transform of $\exp (t \, \M)$ are decreasing with $s$ and may converge only for $\Re (s) > \lambda$. 
As a consequence, if for $s = 0$ the Laplace transform of $\exp (t \, \M)$ diverges then, necessarily, $\lambda \geq 0$. 

Now, separating the diagonal part of $\mathbb{M} - s \mathbbm{1}$, $\mathbb{M}^\text{diag} - s \mathbbm{1}$, to its off-diagonal part, $\mathbb{M}^\text{off}$, and using the Trotter product formula,
\begin{equation}
	e^{t \left( \mathbb{M} - s \mathbbm{1} \right)}
	=
	\lim_{n \to + \infty} 
	\left( 
	e^{\frac{t}{n} \left( \mathbb{M}^\text{diag} - s \mathbbm{1} \right)}
	\,
	e^{\frac{t}{n} \mathbb{M}^\text{off} }
	\right)^n, 
\end{equation}
in the integral in Eq.~\eqref{eq:def_resolvent} yields the path decomposition of the $(i, \, j)$-th entry of the Laplace transform: 
\begin{equation}
	\label{eq:resolvent_path}
	R (s)_{i, \, j} = 
	\sum_{\ell \geq 0 } ~ \sum_{x_1 \cdots x_{\ell}} 
	\delta_{i, \, x_1} \, \delta_{j, \, x_\ell}
	\left( \prod_{k=1}^{\ell-1} w (s)_{x_{k+1} \to x_{k}} \right) \times  \dfrac{1}{| M_{j, \, j} | + s}. 
\end{equation}
In Eq.~\eqref{eq:resolvent_path}, the summations run over all the paths
of arbitrary length $\ell \geq 0$ connecting $i$ and $j$, $i = x_1 \to x_2 \to \cdots \to x_{\ell-1} \to x_\ell = j $; and the elementary weights, $ w (s)_{x \to y}$ for $x \neq y$, are:
\begin{equation}
	w (s)_{x \to y}
	= 
	\dfrac{M_{y, \, x}}{- M_{x, \, x} + s}.
\end{equation}
The latter converges to the elementary transition weights (Eq.~(4) of the main text) as $s \to 0$: 
$w (s)_{x \to y} \longrightarrow w_{x \to y}$. 
The diagonal elements of $R (s)$ can be written as,
\begin{equation}
	R (s)_{i, \, i}
	= 
	\left(  \sum_{\ell \geq 0 } \mathcal{W}(i \to i)(s)^\ell \right) \times  \dfrac{1}{| M_{i,i} | + s},
\end{equation}
in which $\mathcal{W}(i \to i)(s)$ corresponds to the total excursion weight associated to $i$ computed with the elementary weights $w(s)$. 
In the limit $s \to 0$, it converges to the total excursion weight $\W (i \to i)$.  
Then, taking the limit $s \to 0$ yields: 
\begin{equation}
	R (0)_{i, \, i}
	= 
	\left(  \sum_{\ell \geq 0 } \mathcal{W}(i \to i)^\ell \right) \times  \dfrac{1}{| M_{i,i} | },
\end{equation}
which converges if, and only if, $\mathcal{W}(i \to i) < 1$.  
On the other hand, if the Laplace transform converges for $s = 0$ then, necessarily, $\lambda < 0$. 
This finally proves the criterion for exponential growth given in Eq.~(8) of the main text. 

\subsection{\label{sec:first_step}First-step method}

For any Jacobian matrix $\mathbb{M}$, the expression of $\W (X_1 \to X_1)$ can be obtained by solving a linear problem:
\begin{equation}
	\label{eq:first_step_method_general}
	\begin{gathered}
		\W(X_1 \to X_1) = \sum_{i \neq 1} w_{j \to i} \, \W(X_i \to X_1), \\[.5em]
		\W(X_j \to X_1) = w_{j \to 1} + \sum_{i \neq 1, \, j} w_{j \to i} \, \W(X_i \to X_1), ~~\text{for}~~ j \neq 1, 
	\end{gathered} 
\end{equation}
where $w_{j \to i}$ stands for the elementary weight of the transition $j \to i$.  
This method is similar to one used to compute mean first passage time in a Markov chain \cite{norris1998}. 
Using the matrix of the elementary weights, $\mathbb{W}$, whose $(i, \, j)$-th entry is $w_{j \to i}$, the system in Eq.~\eqref{eq:first_step_method_general} can be written as:
\begin{equation}
	\label{eq:first_step_method_matrix}
	\mathbb{T} \cdot \boldsymbol{u}= \mathbf{w}_{\diamond \to 1},
\end{equation}
in such a way that the $i$-th entry of $\boldsymbol{u}$ is $\W(X_i \to X_1)$ and the matrix $\mathbb{T} = \left\lbrace T_{i, \ j} \right\rbrace $ is obtained by setting all the off-diagonal entries in the first column of $- \mathbb{W}^\top$ to zero:
\begin{align}
	T_{1, \, 1} = 1,
	& \quad & 
	T_{i, \, 1} = 0 ~~\text{for}~~i>1, 
	& \quad & 
 	T_{i, \, j} = - W_{j, \, i} = - w_{i \to j} ~~\text{for}~~j>1.
\end{align}
Finally, the entries of $\mathbf{w}_{\diamond \to 1}$ are 
\begin{align}
	\left( \mathbf{w}_{\diamond \to 1}\right)_1 = 0,
	& \quad & 
	\left( \mathbf{w}_{\diamond \to 1}\right)_i = W_{i, \, 1} =  w_{i \to 1}, ~~\text{for}~~ i>1. 
\end{align}
As a result, the total excursion weight, \textit{i.e.}, $u_1 = \W (X_1 \to X_1)$, is expressed as the ratio of two determinants using Cramer's formula: 
\begin{equation}
	\W (X_1 \to X_1) = \dfrac{
			\mathrm{det} \left( \mathbf{w}_{\diamond \to 1} ; \, \mathbb{T}_{\setminus 1} \right)
		}{
		 	\mathrm{det} \left( \mathbb{T} \right)
		},
\end{equation}
where $\left( \mathbf{w}_{\diamond \to 1} ; \, \mathbb{T}_{\setminus 1} \right)$ is the matrix formed by replacing the first column of $\mathbb{T}$ by $\mathbf{w}_{\diamond \to 1} $. 
The sole knowledge of the jacobian matrix suffices to construct $\mathbb{W}$ from which $\mathbb{T}$ and $\mathbf{w}_{\diamond \to 1}$ follow hence, this method provides a way to compute the total excursion weight for any given jacobian matrix. 

Finally, as far as the total excursion weight is concerned, only the first solution of Eq.~\eqref{eq:first_step_method_general} is needed. 

\subsection{\label{sec:proof_single_cycle}Details for the unicyclic autocatalytic cycle}

Defining the following family of determinants~:
\begin{equation}
	\label{eq:Determinant_def}
	\Delta_i = 
	\begin{vNiceArray}{w{c}{1.5cm}w{c}{1cm}w{c}{1cm}w{c}{1cm}w{c}{1cm}w{c}{1.5cm}}[first-col] 
		\X_i 	 &      1  	    & - s_i^+ 			   & 	   0      			  & \Cdots			 	 & 					& 		0		\\[1em]
		\X_{i+1} & - s_{i+1}^- &    1     			   & - s_{i+1}^+ 			  & \Ddots[shorten=0cm]  &  				& \Vdots		\\[1em]
		& 		0		& \Ddots[shorten=0.5cm]  & \Ddots[shorten=0cm]	  & \Ddots[shorten=0.5cm]	 &   			    & 				\\[1em]
		& \Vdots		& \Ddots[shorten=1cm]		   	   & 			  			  &      		 		 &  				& 		0		\\[1em]
		\X_{N-1} & 				& 		          	   & 			  			  & - s_{N-1}^-  	     &  1				& - s_{N-1}^+   \\[1em]			
		\X_N & 		0		& \Cdots	     	   & 			  			  &       0              &  - s_{N}^- 		& 		1 	 
	\end{vNiceArray}.
\end{equation}
for $2 \leq i \leq N$. 
This includes, as a particular case, the determinants $\Delta_2$ and $\Delta_3$ in the total excursion weight given in Eq.~(27) of the main text. 
All the determinants $\Delta_i$ are related to each other by the recursive relation~:
\begin{align}
	\label{eq:recursion_delta}
	2 \leq i \leq N-2 \, , ~~ \Delta_i = \Delta_{i+1} - s_i^+ \, s_{i+1}^- \, \Delta_{i+2}. 
\end{align}
Further, $ \Delta_{N - 1} = 1 -  s_N^- s_{N-1}^+$ and $\Delta_N = 1 $.
An important property which follows from Eq.~\eqref{eq:propensiities_reversible} 
is that for all $i$ :
\begin{equation}
	\label{eq:inequality_propensities}
	s_i^- + s_i^+ \leq 1,
\end{equation}
an inequality which is saturated when $\kappa_i = 0$. 
Since the matrix whose determinant is $\Delta_i$ is diagonally dominated,
we can apply the Gershgorin circle theorem. It follows from this theorem that all  eigenvalues of the matrix lie in at least one Gershgorin circle. These circles are centered at 1 and have a radius, which is either $|s_i^+|$, $|s_N^-|$ or $|s_i^+ + s_i^- |$. 
Since in all these cases, the radius are less than one, 
it follows that, for all $i$, $\Delta_i \ge 0$.
Now, using this property together with Eq. \eqref{eq:recursion_delta}, we obtain :
\begin{equation}
	\label{eq:bound_delta_1}
	\Delta_i - s_i^- \Delta_{i+1} \leq \Delta_{i-1} \leq \Delta_i \leq 1,
\end{equation}
and, combining the recursion relation Eq.~\eqref{eq:recursion_delta} and the inequality Eq.~\eqref{eq:inequality_propensities} yields~:
\begin{align}
	\label{eq:relation_w_1} \Delta_i - s_i^- \Delta_{i+1} & \geq s^+_i \left( \Delta_{i+1} - s_{i+1}^- \Delta_{i+2} \right) \\ 
	\label{eq:relation_w_2} & \geq \prod_{k = i}^{N} s^+_k. 
\end{align}
Note that in Eqs.~\eqref{eq:relation_w_1}-\eqref{eq:relation_w_2}, equality is also reached when degradation vanishes. 
The last inequality implies that 
\begin{equation}
	\dfrac{s_2^- \Delta_3 + \prod_{k=2}^N s_k^+}{\Delta_2} \leq 1,
\end{equation}
and 
\begin{equation}
	\dfrac{\prod_{k=2}^N s_k^+}{\Delta_2} \leq 1.
\end{equation}

Taken together, the above inequalities imply~:
\begin{align}
	\W (X_1 \to X_1) &  = s_1^+ \left( \dfrac{s_2^- \Delta_3 + \prod_{k=2}^N s_k^+}{\Delta_2} + (\mathfrak{p} - 1) \,  \dfrac{\prod_{k=2}^N s_k^+}{\Delta_2} \right) 
	\leq  \mathfrak{p} \, s_1^+ \leq \mathfrak{p}.
\end{align}

\bibliographystyle{apsrev4-2}	
\bibliography{ref.bib}

\end{document}